\documentclass[preprints,article,accept,moreauthors,pdftex]{Definitions/mdpi} 

\firstpage{1} 
\pubvolume{1}
\issuenum{1}
\articlenumber{0}
\pubyear{2021}
\copyrightyear{2020}
\datereceived{} 
\dateaccepted{} 
\datepublished{} 
\hreflink{https://doi.org/} 

\usepackage{bm,amssymb}
\usepackage{xfrac}
\usepackage{color}

\definecolor{purpura}{rgb}{0.5, 0.0, 0.5}
\definecolor{azul}{rgb}{0, 0.0, 0.6}
\definecolor{rojo}{rgb}{0.6, 0, 0}
\definecolor{verde}{rgb}{0, 0.6, 0}
\definecolor{turquesa}{rgb}{0, 0.5, 0.5}
\definecolor{marron}{rgb}{0.6, 0.4, 0}
\definecolor{gris}{rgb}{0.4, 0.4, 0.4}
\definecolor{celeste}{rgb}{0.5, 0.5, 0.8}

\Title{Inferring a property of a large system from a small number of samples}

\TitleCitation{Inferring a property of a large system from a small number of samples}

\Author{Dami\'{a}n G. Hern\'{a}ndez $^{1}$ and In\'{e}s Samengo $^{1}$}

\AuthorNames{Dami\'{a}n G. Hern\'{a}ndez and In\'{e}s Samengo}

\AuthorCitation{Hern\'{a}ndez, D.G.; Samengo, I.}

\address{%
$^{1}$ \quad Department of Medical Physics, Centro Atómico Bariloche and Instituto Balseiro, 8400 San Carlos de Bariloche, Argentina}

\abstract{Inferring the value of a property of a large stochastic system is a difficult task when the number of samples is insufficient to reliably estimate the probability distribution. The Bayesian estimator of the property of interest requires the knowledge of the prior distribution, and in many situations, it is not clear which prior should be used. Several estimators have been developed so far, in which the proposed prior was individually tailored for each property of interest; such is the case, for example, for the entropy, the amount of mutual information, or the correlation between pairs of variables. In this paper we propose a general framework to select priors, valid for arbitrary properties. We first demonstrate that only certain aspects of the prior distribution actually affect the inference process. We then expand the sought prior as a linear combination of a one-dimensional family of indexed priors, each of which is obtained through a maximum entropy approach with constrained mean value of the property under study. In many cases of interest, only one or very few components of the expansion turn out to contribute to the Bayesian estimator, so it is often valid to only keep a single component. The relevant component is selected by the data, so no handcrafted priors are required. We test the performance of this approximation with a few paradigmatic examples, and show that it performs well in comparison to the ad-hoc methods previously proposed in the literature. Our method highlights the connection between Bayesian inference and equilibrium statistical mechanics, since the most relevant component of the expansion can be argued to be the one with the right temperature.}


\begin{document}

\section{Introduction}

Systems with a large number of states appear ubiquitously in the era of big data. Often, the number of available samples is smaller than the number of states, and sometimes, considerably smaller. Although in these circumstances estimating the full probability distribution of the system is highly unreliable, here we show that even in the severe undersampled regime, fairly accurate inferences about the system are possible. The key is that if we are interested in one or a few properties of the system, the complete description of the whole probability distribution is unnecessary. Examples of such properties may be, for instance, the information processing capacity, the correlation between pairs of variables, or the diversity of reachable states in the system.

Let $k$ be the number of available states of the system\footnote{For simplicity, in this work we restrict ourselves to the case of discrete systems. In continuous system with a defined metric, we could possibly overcome the fact that we have few samples by grouping nearby states or reducing the dimension of the space.}, and $\bm{q} = (q_1, \dots, q_k)$ the (unknown) probability distribution that governs the occupation of the states. We assume the number of available samples is $n$, with $n \gtrsim \sqrt{\exp[H(\bm{q})]}$, where $H(\bm{q})$ is the entropy of the system, and $\exp[H(\bm{q})]$ is the effective number of states. For even smaller samples, most of the states remain unsampled, and the few states that are sampled get no more than a single sample. When $n$ reaches the order of $\sqrt{\exp[H(\bm{q})]}$, the probability that a small fraction of the states get more than a single sample becomes large. We are interested in estimating a property $F(\bm{q}) = F(q_1, \dots, q_k)$ whose functional form in terms of $\bm{q}$ is known. The na\"ive approach -- which we here discourage -- would be to first estimate $\bm{q}$, and then evaluate $F(\bm{q})$ in the estimated value. This strategy is called the ``plug-in'' estimator of the property. Although it may work for properties that depend linearly on the probabilities, it is often appallingly biased in nonlinear cases, in particular, for information-related properties that typically bear a logarithmic dependence on the $q_i$. To overcome this problem, some idiosyncratic solutions have emerged during the last decades, valid for particular properties, such as entropy \citep{kraskov2004estimating, nemenman2004entropy, archer2014bayesian, chao2013entropy, grassberger2003entropy} and mutual information \citep{archer2013bayesian, hernandez2019estimating}.

\begin{figure}[H]
\includegraphics[width=10.5 cm]{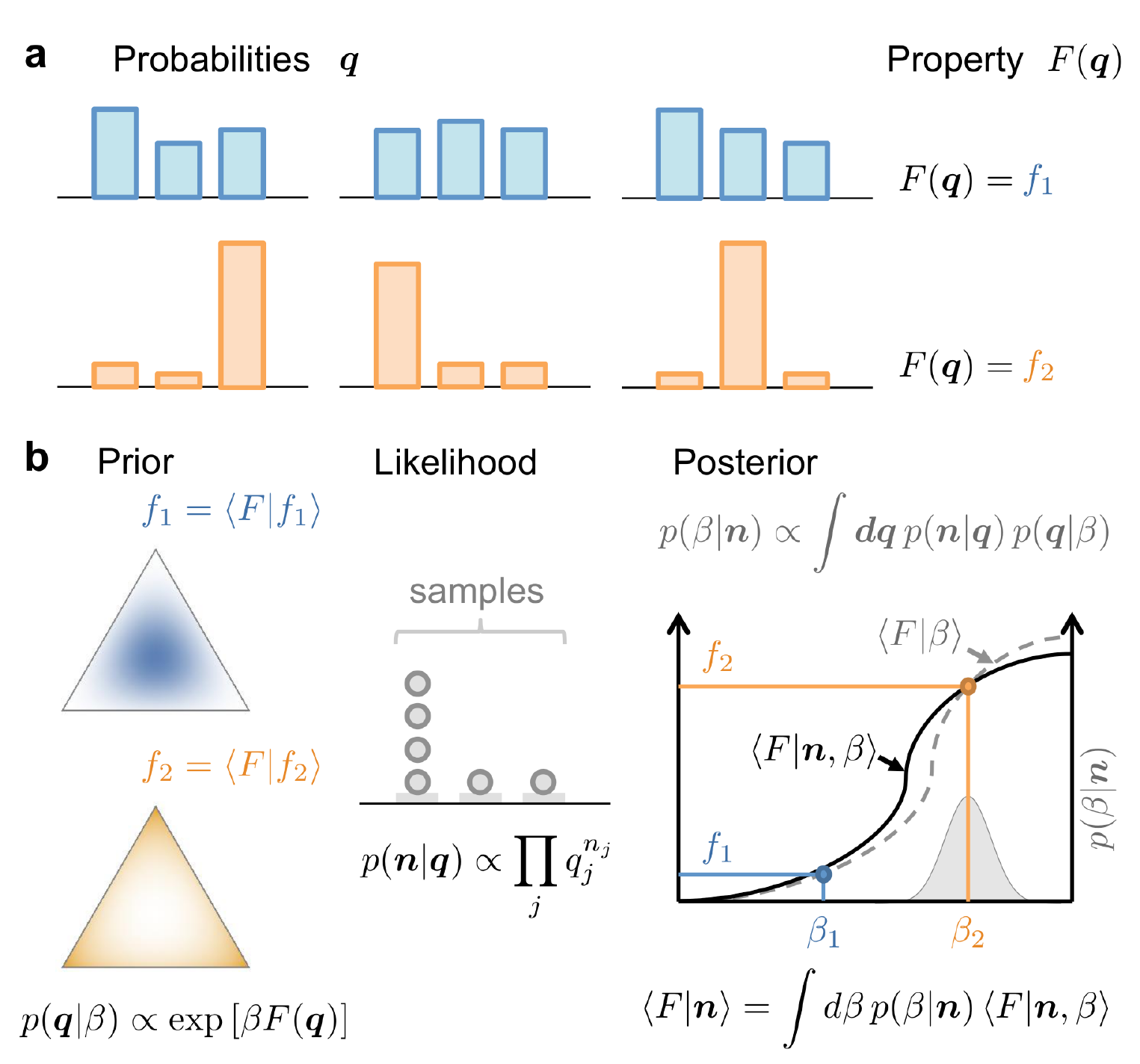}
\caption{Conceptual framework used to estimate the property $F(\bm{q})$ of a stochastic system with probabilities $\bm{q}$ from limited samples $\bm{n}$. (\textbf{a}) Several probabilities $\bm{q}$ can produce the same value of the property. All such $\bm{q}$-vectors belong to the same level surface of $F(\bm{q})$. (\textbf{b}) Left: Example case in which the property $F(\bm{q})$ has circular level surfaces on the simplex embedded in $3$-dimensional space. Two members of the family of base functions used to expand the prior are shown, containing the different $\bm{q}$ vectors shown in (a). These two members are solutions of the Maxentropy problem with two different mean values $f$ of the property. Middle: The histogram $\bm{n}$ generated by the sampled data produces a likelihood $p(\bm{n}|\bm{q})$ that selective favours a specific range of surface levels. Right: The most favoured $f$ value (or equivalently, $\beta$ value) is the one for which the prior and posterior estimates of the property coincide.\label{fig:fig1}}
\end{figure}

Here our goal is to develop a principled way to approach this problem for properties $F(\bm{q})$ that may be arbitrary, as long as they vary smoothly with $\bm{q}$. The clue relies on the fact that the simplex ${\cal Q}$ containing the set of possible $\bm{q}$-vectors can be foliated by the level surfaces of $F(\bm{q})$. If only the value of $F$ matters, estimating the full $\bm{q}$ is an overkill; the only inference that needs to be made is about the level surface to which $\bm{q}$ belongs. This is a one-dimensional inference problem, and hence, much simpler than the $k$-dimensional problem of estimating $\bm{q}$. We here estimate $F$ within a Bayesian framework, using a prior distribution that weighs the different level surfaces according to a MaxEntropy principle.

\section{Rationale of the inference process }

\subsection{A Bayesian approach to the inference problem}
Our goal is to estimate a property $F(\bm{q})$ of a stochastic system composed by a large number $k$ of states, when we only have access to $n$ samples, with $n \gtrsim \sqrt{\exp[H(\bm{q})]}$. The probabilities $\bm{q} = (q_1, \dots, q_k)$ constitute a complete description of the system (see Fig.~\ref{fig:fig1}). One of our main assumptions is that the number of effective states is large, that is, $\mathrm{exp}[H(\bm{q})] \gg 1$, where $H(\bm{q})$ is the entropy. Moreover, this number is also assumed to be larger or in the order of magnitude of the number of samples $n$. 

We choose to work with the estimator $\hat{F} = \langle F(\bm{q} | \bm{n}) \rangle$ defined as the mean value of the property $F(\bm{q})$ when conditioned by the measured data $\bm{n}$. The average $\langle \cdot \rangle$ is calculated with the posterior distribution $p(\bm{q}|\bm{n})$, and the $k$-dimensional vector $\bm{n}$ has components $n_1, \dots, n_k$, with $n_i$ equal to the number of samples in which the system was observed in the $i$-th state. Therefore,
\begin{linenomath*}
\begin{equation}
    \displaystyle    \hat{F}=\langle F|\bm{n}\rangle = \int \bm{dq}\, F(\bm{q})\,p(\bm{q}|\bm{n}).
    \label{eq01:meanFpqn}
\end{equation}
\end{linenomath*}
Throughout the paper, all integrals in $\bm{q}$ should be restricted to the space where $\bm{q}$ is defined, typically, the simplex embedded in $\mathbb{R}^k$, or the cartesian product of several simplexes, in the multivariate case. The estimator of Eq.~\ref{eq01:meanFpqn} minimizes the mean square error of the estimation \citep{lehmann1998}. Using Bayes' rule, the posterior can be written in terms of the likelihood $p(\bm{n}|\bm{q})$ and the prior $p(\bm{q})$, so that
\begin{linenomath*}
\begin{equation}
    \displaystyle    \langle F|\bm{n}\rangle = \int \bm{dq}\, F(\bm{q})\,\frac{p(\bm{n}|\bm{q})p(\bm{q})}{p(\bm{n})}.
    \label{eq02:meanFpqn}
\end{equation}
\end{linenomath*}
The likelihood $p(\bm{n}|\bm{q})$ is the only factor that depends on the sampled data $\bm{n}$, and for discrete states, it is a multinomial distribution, namely 
\begin{linenomath*}
\[
p(\bm{n}|\bm{q}) \propto \prod_j q_j^{n_j}.
\]
\end{linenomath*}

The only factor that still needs to be defined is the prior $p(\bm{q})$, and the strategy underlying this choice is the topic of this paper. We want the prior to produce an inductive bias that acknowledges our ignorance about $F(\bm{q})$. Such bias will be useful to overcome the scarcity of samples, but it will only be suitable to estimate the specific $F(\bm{q})$ of the problem at hand, and not others. 

\subsection{Defining the prior on the level surfaces of the property}

\label{sec:dospuntodos}

The search for the prior should consider all candidate $p(\bm{q})$ that can be defined on the simplex embedded in $\mathbb{R}^k$. However, in this subsection we show that the search can be reduced to a subset, since the variations of $p(\bm{q})$ inside the level surfaces of $F(\bm{q})$ are irrelevant, in the sense that they have no bearing on the value of the estimator $\hat{F}$. To see this point, we introduce a change of variables $\bm{q} \to \bm{q}'$ to perform the integral of Eq. \ref{eq02:meanFpqn}. Specifically, we define the first two coordinates of the new variables
\begin{linenomath*}
\begin{eqnarray}
q_1'(\bm{q}) &=& f = F(\bm{q}) = \text{property} \\
q_2'(\bm{q}) &=& \ell = q(\bm{n} | \bm{q}) = \text{likelihood }, \nonumber 
\end{eqnarray}
\end{linenomath*}
and the remaining coordinates $q_3'(\bm{q}), \dots, q_k'(\bm{q})$ may be chosen arbitrarily, as long as the transformation $\bm{q} \to \bm{q}'$ be invertible. In the new variables, 
\begin{linenomath*}
\begin{eqnarray}
    \displaystyle    \langle F|\bm{n}\rangle &=&  \int {\rm d}f \ f \ \int {\rm d}\ell \ \ell \int {\rm d}q_3' \dots {\rm d}q_k' \  \frac{p(\bm{q}')}{p(\bm{n})} \ \left| \frac{\partial{q_1, \dots, q_k}}{q_1', \dots, q_k'} \right| \nonumber \\
    &=&\int {\rm d}f \ f \ \int {\rm d}\ell \ \ell \ p(f, \ell | \bm{n}), \ \ \ \ \text{with} \nonumber \\
    p(f, \ell | \bm{n}) &=& \int {\rm d}q_3' \dots {\rm d}q_k' \  \frac{p(\bm{q}')}{p(\bm{n})} \ \left| \frac{\partial{q_1, \dots, q_k}}{q_1', \dots, q_k'} \right| \nonumber \\
    &=& \int {\rm d}\bm{q} \ \frac{p(\bm{q})}{p(n)} \ \delta[f(\bm{q}) - f] \ \delta[L(\bm{q}) - \ell].   \label{eq03:primedvariables} 
\end{eqnarray}
\end{linenomath*}
Equation \ref{eq03:primedvariables} shows that it does not matter how the prior $p(\bm{q})$ distributes the density inside the manifold ${\cal M}_{f\ell}$ obtained by intersecting the level surface $f$ of the property $F$ with a level surface $\ell$ of the likelihood $L$. Only the integral of $\bm{q}$ inside ${\cal M}_{f\ell}$ has an effect on $\hat{F}$. Therefore, we opt to only consider priors $p(\bm{q})$ that are constant inside ${\cal M}_{f\ell}$, since for any prior not constant in ${\cal M}_{f\ell}$, a prior that is constant in ${\cal M}_{f\ell}$ exists that produces the same estimation $\hat{F}$. Our task is therefore to design how the prior $p(\bm{q})$ changes with $f$ and $\ell$. A crucial point is that the level surfaces of $L(\bm{q})$ vary with the data $\bm{n}$. Yet, by definition, a prior cannot depend on the measured data. Therefore, we only work with priors that, when written in the $\bm{q}$-coordinates, only depend on $\bm{q}$ through $F(\bm{q})$. In other words, we assume that a function $g$ exists, such that $p(\bm{q}) = g[F(\bm{q})]$. By limiting the search to priors of this type, for each value of $f$, we impose maximal uncertainty about $\bm{q}$, as dictated by a maxentropy principle. 

It should be noted that the assumption $p(\bm{q}) = g[F(\bm{q})]$ does not imply that, when written in the $\bm{q}'$-coordinates the  prior $p(\bm{q}')$ be independent of $\ell = q_2'$, since the jacobian of the transformation of variables may well depend on $\ell$. Moreover, the marginal prior $p(f, \ell | \bm{n})$ may also depend on $\ell$, since the level surface in which  Eq.~\ref{eq03:primedvariables} is calculated depends on $L(\bm{q})$. Therefore, $p(f, \ell | \bm{n})$ is allowed to depend on $\ell$, but only due to the geometrical structure that the Delta function in $L(\bm{q})$ restricts the integration region of Eq.~\ref{eq03:primedvariables}.

\subsection{Decomposing the prior as a linear combination of a family of base priors}

The set of priors $p({\bm q})$ for which a function $g: \mathbb{R} \to \mathbb{R}$ exists such that $p(\bm{q}) = g[F(\bm{q})]$ is equal to all the distributions of the form
\begin{linenomath*}
\begin{equation} \label{eq:deltas}
p(\bm{q}) = \int {\rm d}f \ g(f) \ \delta[F(\bm{q}) - f]
\end{equation}
\end{linenomath*}
that can be obtained by varying $g(f)$. Therefore, we have reduced the problem of specifying $p(\bm{q})$ to the problem of specifying $g(f)$, which is a drastically simpler object, given that $\bm{q}$ has $(k-1)$ independent dimensions, whereas $f$ has only a single one.

We now show that, when the number of samples is sufficiently large, the determination of $g$ becomes unnecessary, since for large $k$, the measured data $\bm{n}$ often enter into the likelihood in such a way that only a small range of $f$-values remains with significantly non-zero probability. Therefore, all the $g$-functions that do not vary drastically within the permitted range give rise to the same estimation. In such cases the data alone dictate the value of the estimator through the likelihood, making the discussion about priors essentially inconsequential. We now prove these statements. Replacing Eq.~\ref{eq:deltas} in \ref{eq02:meanFpqn}, the estimator becomes
\begin{linenomath*}
\begin{eqnarray}
    \langle F | \bm{n} \rangle &=& \frac{1}{p(\bm{n})} \ \int {\rm d}\bm{q} \ p(\bm{n} | \bm{q}) \ F(\bm{q}) \ g[F(\bm{q})] \ \delta[F(\bm{q}) - f] \nonumber \\
    &=&  \int {\rm d}f \ f \ p(f|\bm{n}) \label{eq:desc1}
\end{eqnarray}
with
\begin{eqnarray}
    p(f|\bm{n}) &=& \frac{g(f)}{p(\bm{n})} \ \int {\rm d}\bm{q} \ p(\bm{n} | \bm{q}) \ \delta[F(\bm{q}) - f] \nonumber \\
    &\propto& \frac{g(f)}{p(\bm{n})} \ \int {\rm d}\bm{q} \ \mathrm{e}^{-n \ D_{\mathrm{KL}}(\frac{1}{n}\bm{n}||\bm{q})}\ \delta[F(\bm{q}) - f], \label{eq:dkl}
\end{eqnarray}
\end{linenomath*}
where $D_{\mathrm{KL}}(\bm{a}||\bm{b})$ is the Kullback-Leibler divergence between the distributions $\bm{a}$ and $\bm{b}$, and the $i$-th component of the vector $\frac{1}{n}\bm{n}$ is $n_i/n$. The  divergence is always non-negative, and it only vanishes when the two distributions coincide \citep{cover2012elements}. Therefore, if the number of samples $n$ is sufficiently large, the result of the integral is significantly different from zero only for level surfaces that pass close enough of the sampled frequencies $\bm{n}/n$, and is maximal for the one that contains the sampled frequencies. The data $\bm{n}$ thereby select the range of $f$ values that are compatible with the observations. The factor $n$ in the exponent of Eq.~\ref{eq:dkl} implies that the allowed range of $f$-values becomes increasingly narrow as the number of samples grows. The crucial point in this reasoning is that for sufficiently narrow ranges, the shape of the prior $g(f)$ becomes irrelevant. If the range is much narrower than the typical scales in which $g(f)$ varies, for all practical matters, $g(f)$ is constant within this range, and has no bearing in the estimation. This is the situation in which we no longer need to bother about the prior.

When that limit is reached, the likelihood becomes a Delta distribution, and the only $\bm{q}$-vector that contributes to a given $f$-value is $\bm{q} = \bm{n}/n$. In this extreme case, when $g(f)$ is uniform, the estimator $\langle F | \bm{n} \rangle$ converges to the plug-in estimator. Yet, the plug-in estimator is only justified when the likelihood indeed reaches the Delta-like behaviour, which only happens for $n \gg k$. Before this limit, the plug-in estimator completely neglects the width of the peak of the likelihood around the sampled frequencies $\bm{n}/n$, and thereby, the possibility that a whole collection of $\bm{q}$-vectors in the vicinity of $\bm{n}/n$ contribute to each $f$. Allowing for this possibility is important, since in the undersampled regime, there is a large degree of uncertainty about the true $\bm{q}$ that generated the data $\bm{n}$. By neglecting this uncertainty, the plug-in estimator produces the woeful biases mentioned before. The obvious solution is not to neglect the width of the likelihood. However, if no approximations are made, it is difficult to calculate or even to estimate $p(f|\bm{n})$ analytically, given that Eq.~\ref{eq:dkl} involves an integral over an arbitrarily shaped manifold.

To overcome this problem, we would like to design another estimator that retain the insensitivity to the prior, but still, give a chance to a collection of $\bm{q}$ vectors to contribute to each $f$. The size of this collection is determined by the width of the likelihood, that is, by the total number of samples $n$, and the obtained frequencies $\bm{n}/n$. The averaged contributions of all the $\bm{q}$ vectors that have non-vanishing likelihoods produces an estimated $\hat{F}$ that may be either smaller or larger than the plug-in value. In other words, all the $\bm{q}$-vectors different from $\bm{n}/n$ may tend to either increase or decrease $F$ as compared with $F(\bm{n}/n)$, and the direction and size of this shift depends on the behaviour of $F(\bm{q})$ around $\bm{q} = \bm{n}/n$. In what follows, we propose an alternative expansion of the prior different from Eq.~\ref{eq:deltas}, that is formulated in terms of a parameter that controls the relative weight of the level surfaces that tend to increase vs. those that tend to decrease $F$.

Using the previous analysis as an inspiration, from now on we consider a narrower set of priors, that is a proper subset of the one defined by Eq.~\ref{eq:deltas}. A restriction in the set of possible priors does not invalidate the analysis of how the width of the likelihood depends on the sampled data, it only reduces the set from which priors are selected. We begin by proposing the decomposition
\begin{linenomath*}
\begin{equation}
        \displaystyle p(\bm{q}) = \int {\rm d}f \, p(\bm{q}|f) \ g(f),
    \label{eq03:familyPrior}
\end{equation}
\end{linenomath*}
for a conveniently chosen family of priors $p(\bm{q}|f)$ that now need not coincide with $\delta[F(\bm{q}) - f]$, but are still assumed to only depend on $\bm{q}$ through the property $F(\bm{q})$. In other words, we assume that a family of normalizable functions $g_f:\mathbb{R} \to \mathbb{R}$ exists, such that
\begin{linenomath*}
\[
    p(\bm{q}|f) = g_f[F(\bm{q})].
\]
\end{linenomath*}
Before continuing, it is important to note that in the new decomposition of Eq.~\ref{eq03:familyPrior}, the parameter $f$ can no longer be identified as exactly the value of the property $F(\bm{q})$ in a given level surface, since the delta functions are no longer present to restrict $p(\bm{q}|f)$ to a single level surface of $F(\bm{q})$. Therefore, from now on, $f$ should be regarded as a formal label that parametrises the set of base functions used in the expansion. Yet, later on we will choose a family of base functions for which there is still a connection between $f$ and $F(\bm{q})$, but the connection will turn out to be probabilistic (see next section). 

By inserting Eq.~\ref{eq03:familyPrior} in Eq.~\ref{eq02:meanFpqn}, the estimator becomes
\begin{linenomath*}
\begin{eqnarray}
    \langle F | \bm{n} \rangle &=& \int {\rm d}q \ F(\bm{q}) \ \frac{p(\bm{n} | \bm{q})}{p(\bm{n})} \ \int {\rm d}f \ p(\bm{q}|f) \ g(f) \nonumber \\
    &=& \int {\rm d}f \ \frac{g(f)}{p(\bm{n})} \  p(\bm{n} | f) \ \int {\rm d}\bm{q} F(\bm{q}) \ p(\bm{n}| \bm{q}) \  \frac{p(\bm{q}|f)}{p(\bm{n}|f)} \nonumber
\end{eqnarray}
\end{linenomath*}
where we have introduced the likelihood of the data for each value of the parameter, 
\begin{linenomath*}
\begin{equation} \label{eq:pedeenedadoefe}
p(\bm{n}|f) = \int {\rm d}\bm{q} \ p(\bm{q} | f) \ p(\bm{n}| \bm{q}).
\end{equation}
\end{linenomath*}
With this definition, the estimator reads
\begin{linenomath*}
\begin{equation}    
\langle F|\bm{n} \rangle = \int {\rm d}f \ p(f | \bm{n}) \ \langle F|\bm{n}, f\rangle, \label{eq:integroenefe}
\end{equation}
\end{linenomath*}
where
\begin{linenomath*}
\begin{equation}
    p(f|\bm{n}) = \frac{p(\bm{n}|f) \ g(f)}{p(\bm{n})} \label{eq04:GuessEvidence}
\end{equation}
\end{linenomath*}
represents the amount of evidence in favour of each $f$-value provided by the measured data $\bf{n}$, and
\begin{linenomath*}
\begin{eqnarray}
    \langle F | \bm{n}, f \rangle &=& \int {\rm d}\bm{q} \ F(\bm{q}) \ \ p(\bm{q}| \bm{n}, f)  \label{eq03:Flambda} \\ &=&  \int {\rm d}\bm{q} \ F(\bm{q}) \ p(\bm{n}| \bm{q}) \  \frac{p(\bm{q}|f)}{p(\bm{n}|f)}. \nonumber
\end{eqnarray}
\end{linenomath*}
is the estimation of the property $F(\bm{q})$ conditional on the measured data $\bm{n}$ and the parameter $f$. 

Equation \ref{eq:integroenefe} is homologous to Eq.~\ref{eq:desc1}, for base priors that are not Delta distributions. Using Bayes' rule (Eq.~\ref{eq04:GuessEvidence}), the evidence $p(f|\bm{n})$ can be written in terms of the marginal likelihood $p(\bm{n}|f)$ of Eq.~\ref{eq:pedeenedadoefe}. This evidence is defined by an integral in $\bm{q}$-space that contains the multinomial likelihood $p(\bm{n}|\bm{q})$ embodying the Kullback-Leibler divergence. Therefore, just as before, the data $\bm{n}$ still select a range of $f$-values, only that know, we cannot identify $f$ as an instantiation of $F$ on one specific region. Still, the value of $f$ that maximises the posterior $p(f|\bm{n})$ is the one that makes the largest contribution to the integral in Eq.~\ref{eq:integroenefe}. Before, only keeping the optimal $f$-value yielded to almost the plug-in estimator (except for the effect of $g(f)$). In the present case, only keeping the optimal $f$-value means to replace the integral in Eq.~\ref{eq:integroenefe} by the evaluation of $\langle F | \bm{n}, f \rangle$ at the value of $f = f^*$ that maximises $p(f|\bm{n})$. This procedure is henceforth denominated the \emph{MAP estimator}, for \emph{Maximum A Posteriori}. With the present decomposition, this procedure does not yield the plug-in estimator, because $\langle F | \bm{n}, f^* \rangle$ contains an integral that sweeps through a whole range of level surfaces. For the family of base priors selected in the following subsection, the MAP estimator performs substantially better than the plug-in estimator, and often, also better than custom-made estimators designed for specific properties. In those cases in which the shape of $g(f)$ can be argued to play no relevant role, the MAP estimator can be replaced by an empirical Bayes estimator, in which the selected $f^*$ value maximises the marginal likelihood $p(\bm{n}|f)$, instead of the marginal posterior $p(f|\bm{n})$.


\subsection{A Maxentropy strategy to select the base of functions to expand the prior}


We must now select the family of distributions $p(\bm{q}|f)$ that serve as a base to expand the prior $p(\bm{q})$, as dictated by Eq.~\ref{eq03:familyPrior}. Before, when the expansion in Delta distributions was used, each element of the base was associated with a single $f$ value, equal to the property $F$ evaluated on the corresponding level surface. A natural relaxation of this condition, while still fulfilling the requirement that the elements $p(\bm{q}|f)$ have the same level surfaces as $F(\bm{q})$, is to demand that $f$ be the \emph{mean} value of $F(\bm{q})$, when the average is weighted by $p(\bm{q}|f)$. If, after imposing this requirement, we insist that the base functions have no additional structure, then the shape of $p(\bm{q}|f)$ can be derived from the Maxentropy principle \citep{jaynes1957a, jaynes1957b}, in which the entropy ${\cal H}[p(\bm{q} | f)]$ is maximised, conditioned to the restriction
\begin{linenomath*}
\begin{equation}
    \int {\rm d}\bm{q} \ F(\bm{q}) \ p(\bm{q}|f) = f.
    \label{eq06:condFprior}
\end{equation}
\end{linenomath*}
The solution of this maximisation problem is
\begin{linenomath*}
\begin{equation}
    \displaystyle p(\bm{q}|f)=\frac{\mathrm{e}^{\beta F(\bm{q})}}{\int {\rm d}\bm{q}' \mathrm{e}^{\beta F(\bm{q}')}}=\frac{\mathrm{e}^{\beta F(\bm{q})}}{\mathcal{Z}_0(\beta)},
    \label{eq07:priorMaxEnt}
\end{equation}
\end{linenomath*}
where the hyperparameter $\beta$ is a function of $f$, that is, of the mean value of the property. The correspondence between each $f$-value and each $\beta$-value implies that both these parameters constitute valid tags to designate a member of the base. For this reason, in what follows we use the two parametrizations interchangeably, depending on what we want to stress. We warn the reader that we pass from one to the other, tagging the elements of the base equivalently as $p(\bm{q}|f)$ or as $p(\bm{q}|\beta)$, understanding that there is a one-to-one mapping between $f$ and $\beta$. In the nomenclature of Amari \citep{amari2000, amari2001}, the parameter  appearing linearly in the exponent of the distribution (for us, $\beta$) is referred to as an \emph{exponential} system of coordinates of the space of parameters. When $f$ is used, the coordinates are called \emph{mixed}. In equilibrium statistical mechanics, for example, $\beta$ is often proportional to the negative of the inverse of the temperature, whereas $f$ is proportional to the mean energy of a state---and of course, the two are related.

In the exponential family of Eq.~\ref{eq07:priorMaxEnt}, all level surfaces of $F(\bm{q})$ contribute to each $p(\bm{q}|f)$, but some are more relevant than others. If $\beta$ is large and positive, $p(\bm{q}|f)$ is dominated by the level surface with maximal $F(\bm{q})$, and rapidly dies out when we depart from it. The mean value of $F(\bm{q})$ weighted by this $p(\bm{q}|\beta \to +\infty)$ is equal to the maximum of $F$ on the simplex. If $\beta = 0$, all level surfaces contribute uniformly. If $\beta$ is large and negative, the level surface with minimal $F(\bm{q})$ has maximal relevance, and the mean value of $F(\bm{q})$ is equal to the minimum of $F$ on the simplex. Therefore, $\beta$ operates as a tuning knob that raises or lowers the relevance of different level surfaces, ranking them by the value of $F$. Just as happened when expanding the prior in a base of Delta functions, the mean value of $F$ shifts from its minimum to its maximum, by adjusting the parameter. But contrasting with that previous case, the expansion in exponential functions allows each element of the base to spread out to different level surfaces, allowing thereby that a whole diversity of surfaces contribute to each $\langle F | \beta \rangle$.

\subsection{The estimation of the property for each member of the base}

The posterior distribution conditioned on a specific member $\beta$ of the base is
\begin{linenomath*}
\begin{eqnarray}
    p(\bm{q}|\bm{n}, \beta) &=& \frac{p(\bm{n} | \bm{q}) \ p(\bm{q} | \beta)}{p(\bm{n} | \beta)} \nonumber \\
    &=&\frac{ \mathrm{e}^{\beta F(\bm{q})} \ \prod_j q_j^{n_j} }{\int {\rm d} \bm{q}' \ \mathrm{e}^{\beta F(\bm{q'})} \ \prod_j {q'}_j^{n_j} } \nonumber \\
    &=& \frac{\mathrm{e}^{\beta F(\bm{q})} \ \prod_j q_j^{n_j} }{\mathcal{Z}(\bm{n},\beta)}.
    \label{eq08:posterior}
\end{eqnarray}
\end{linenomath*}
Replacing this expression in Eq.~\ref{eq03:Flambda}
\begin{linenomath*}
\[
\langle F|\bm{n}, \beta \rangle = \frac{1}{\mathcal{Z}(\bm{n}, \beta)} \ \int {\rm d}\bm{q} \ F(\bm{q}) \ \mathrm{e}^{\beta F(\bm{q})} \ \prod_j q_j^{n_j} ,
\]
where the posterior partition function $\mathcal{Z}(\bm{n}, \beta)$ is calculated with the product of the prior distribution $p(\bm{q}|\beta)$ and the likelihood $p(\bm{n}|\bm{q})$.
\end{linenomath*}
Just as in statistical mechanics \citep{jaynes1957a}, this integral is equal to 
\begin{linenomath*}
\begin{equation}
\langle F|\bm{n}, \beta \rangle = \partial_\beta \ \log \mathcal{Z}(\bm{n}, \beta). \label{eq09:memberPostGuess}
\end{equation}
\end{linenomath*}
Moreover, following the same reasoning used in statistical mechanics, the variance of the estimation reads
\begin{linenomath*}
\[
\langle \Delta F^2|\bm{n}, \beta \rangle = \partial^2_{\beta}\log\mathcal{Z}(\bm{n}, \beta).
\]
\end{linenomath*}
Importantly, in several applications the partition function $\mathcal{Z}(\bm{n}, \beta)$ can be calculated analytically. In those cases where an analytical treatment is impossible, the shape of $\mathcal{Z}(\bm{n}, \beta)$ can often be well approximated by one-dimensional numerical integrals, which still enables the computation of $\langle F | \bm{n}, \beta \rangle$.

\subsection{Considering different $\beta$ values}

Equation~\ref{eq:integroenefe} states that the full estimation $\langle F|\bm{n} \rangle$ results from integrating $\langle F|\bm{n}, \beta \rangle$ in $\beta$, each contribution weighted by the evidence $p(\beta | \bm{n})$. So we now analyse this evidence.

When $p(\bm{q}|\beta)$ is given by   Eq.~(\ref{eq07:priorMaxEnt}), the evidence of Eq.~(\ref{eq04:GuessEvidence}) becomes
\begin{linenomath*}
\begin{equation}
    \displaystyle \log p(\beta|\bm{n}) \propto \log p(\beta)+ \log p(\bm{n}|\beta)=\log p(\beta)+\log\left[ \frac{\mathcal{Z}(\bm{n},\beta)}{\mathcal{Z}_0(\beta)}\right],
    \label{eq10:posteriorEvidence}
\end{equation}
\end{linenomath*}
where the constant of proportionality does not depend on $\beta$. 

The first term of Eq.~\ref{eq10:posteriorEvidence} contains
the prior expectations on $\beta$. The second one is the marginal likelihood $p(\bm{n}|\beta)$, and describes the way in which the data $\bm{n}$ favour a specific range of level surfaces. It is equal to the ratio of the posterior and the prior partition functions.  The $\beta$ value that makes the largest contribution to this ratio, henceforth called $\beta_0$, can be found by taking the derivative of the logarithm  of the ratio, and set it to zero,
\begin{linenomath*}
\begin{equation}
    \displaystyle \partial_{\beta}\log p(\bm{n}|\beta)\bigg\rvert_{\beta_0}= \langle F|\bm{n}, \beta_0\rangle-\langle F|\beta_0\rangle =0.
    \label{eq11:peakMargLik}
\end{equation}
\end{linenomath*}
Therefore, $\beta_0$ is the parameter for which the prior and posterior estimations of the property coincide. This idea is illustrated in the last panel in Fig.~\ref{fig:fig1}. 


\section{Applications to different properties}
\label{sect:examples}

In this section, we apply the method to estimate the marginal probability of a bivariate distribution (Sect.~\ref{sect:marginal}), the amount of mutual information (Sect.~\ref{sec:mi}), and the entropy (Sect.~\ref{sec:entropia}). All these quantities are examples of properties $F(\bm{q})$ that have been extensively discussed in the literature \citep{panzeri1996analytical, samengo2002estimating, paninski2003estimation, nemenman2004entropy, archer2013bayesian}. Yet, as far as we know, all the methods developed so far---at least, the ones that perform decently in the severe undersampled regime---are idionsynchratically tailored to each specific property. Since the method proposed here applies to any property, we test it here to compare its performance to the more handcrafted approaches. 

\subsection{Marginal probability}
\label{sect:marginal}
As a first example, we consider a system which states are tagged by two labels: $x$, with many equiprobable states and $y\in\{0,1\}$, with just two. The property we are interested in estimating is the marginal probability of $y=1$, denoted as $q_1$, defined as
\begin{linenomath*}
\[
q_1(\bm{q}) = \sum_x q_x \, q_{1|x} = \frac{1}{k_x} \sum_x q_{1|x} \ ,
\]
\end{linenomath*}
were $k_x$ is the number of $x$-values. Since we assume all $x$-values have the same probability, here $\bm{q}$ represents the collection of conditional probabilities $\{q_{1|x}\}_{x=1}^k$. This case constitutes a toy example, since the plug-in estimator $\hat{q}_1 = n_1/n$ is unbiased and efficient, irrespective of the value of $n_x$. Still, we discuss this case to exemplify the procedure described in the previous section.

According to the procedure developed in the previous section, the base functions to be used to expand the prior are solutions of a Maxentropy problem conditioned to have $q_1(\bm{q}) = f$, which in exponential coordinates reads (see Eq.~\ref{eq07:priorMaxEnt}),
\begin{linenomath*}
\begin{equation}
    \displaystyle p(\bm{q}|\beta) \propto \exp\left(\frac{\beta}{k_x} \sum_x q_{1|x}\right) \propto \left(\prod_x \mathrm{e}^{q_{1|x}} \right)^{\beta/k_x} \,\propto\, \prod_x p(q_{1|x}|\beta).
    \label{eq12:priorq1}
\end{equation}
\end{linenomath*}
For each $\beta$, this solution factorises into independent terms. The exponential function favours either small or large values of $q_{1|x}$, depending on the sign of $\beta$. As the likelihood also factorises $p(\bm{n}|\bm{q}) \propto \prod_x q_{1|x}^{n_{1x}}(1-q_{1|x})^{n_{0x}}$, so does the posterior,
\begin{linenomath*}
\begin{equation}
    \displaystyle p(q_{1|x}|n_{1x},n_{0x},\beta)= \frac{q_{1|x}^{n_{1x}}(1-q_{1|x})^{n_{0x}}\, \mathrm{e}^{\beta q_{1|x}/k_x} }{M(n_{1x}+1,n_x+2,\beta/k_x)},
    \label{eq13:posteriorq1x}
\end{equation}
\end{linenomath*}
where $M(a,b,z)$ is the confluent hypergeometric function \cite{abramowitz1988handbook}. Then the partition function corresponds to $\mathcal{Z}(\mathbf{n}, \beta)=\prod_x M(n_{1x}+1,n_x+2,\beta/k_x)$. The prior partition function $\mathcal{Z}_0(\beta)$ can always be obtained from $\mathcal{Z}(\mathbf{n}, \beta)$ by setting $\bm{n} = \bm{0}$. 

In Eq.~\ref{eq13:posteriorq1x}, the exponential factor is inherited from the prior $p(q_{1|x}|\beta)$. In general, in the method proposed here each $\beta$ biases the estimation towards the corresponding $f$ value, in order to get an unbiased estimate of the property when all the $\beta$ values are considered.

With the partition function, we derive the estimate for the property $q_1$ of each member $\beta$ of the family as 
\begin{linenomath*}
\begin{equation}
    \displaystyle \langle q_1|\bm{n},\beta \rangle = \frac{1}{k_x}\sum_x \langle q_{1|x}|n_{1x},n_{0x},\beta \rangle= \frac{1}{k_x}\sum_x \frac{(n_{1x}+1)}{(n_x+2)}\frac{M(n_{1x}+2,n_x+3,\beta/k_x)}{M(n_{1x}+1,n_x+2,\beta/k_x)}.
    \label{eq14:guessq1}
\end{equation}
\end{linenomath*}

The parameter $\beta_0$ that has a maximal contribution to the estimation is the one for which $\langle q_1|\bm{n},\beta_0 \rangle = \langle q_1|\beta_0 \rangle$, as established in Eq.~(\ref{eq11:peakMargLik}). This equations is often difficult to solve analytically. However, in the present case it can be easily solved in the extreme undersampled regime ($n \ll k_x$) with no coincidences ($n_x \le 1$), where there is a fraction $f_1=n_1/n$ of $x$ states with $y=1$ and a fraction $(1-f_1)$ of states with $y=0$. This condition translates into
\begin{linenomath*}
\begin{equation}
    \displaystyle (1-f_1)\frac{M(2,4,\gamma)}{3 M(1,3,\gamma)}+f_1\frac{2 M(3,4,\gamma)}{3 M(2,3,\gamma)}=\frac{M(2,3,\gamma)}{2 M(1,2,\gamma)}=\langle q_1|\bm{n},k_x \, \gamma \rangle,
    \label{eq15:solq1peak}
\end{equation}
\end{linenomath*}
where $\gamma=\beta/k_x$. Taking into account the contiguous relations of the confluent hypergeometric function, this equation leads to $\langle q_1|\bm{n},\beta_0 \rangle= f_1=n_1/n$, which is the na\"{\i}ve, plug-in, unbiased estimator. Usually for other properties, the plug-in estimator provides quite appalling results in the severe undersampled regime, and there is no clear path to a low bias estimator. We believe that the proposed approach, focused on priors defined on the level surfaces, can be useful in these cases. In the two following subsections we analyse other more challenging properties such as the mutual information and the entropy.

\subsection{Mutual information}
\label{sec:mi}
The estimation of the amount of mutual information in a severely undersampled discrete system is a difficult task \citep{kraskov2004estimating, montemurro2007tight, kolchinsky2017estimating, belghazi2018mutual, safaai2018information, holmes2019estimation}. Several of the approaches developed previously \citep{nemenman2004entropy, archer2013bayesian} are based on estimators for the total and conditional entropies. In addition, in a previous publication we also developed a method that was focused on the dispersion of the conditional probabilities in relation to the marginal \citep{hernandez2019estimating}. 

We work with the same system introduced in the previous subsection. We assume $k_x\gg 1$, and all $x$-values have with equal probability $q_x=1/k_x, \ \forall x$. The two states for the label $y\in\{0,1\}$ are assumed to have equal marginal probabilites, $q_y = \sfrac{1}{2}, \ \forall y$. For this system, we want to find the mutual information $I = I(X; Y)$ namely,
\begin{linenomath*}
\begin{equation}
    \begin{array}{rl}
        I &\displaystyle= \sum_{xy} q_x\, q_{y|x}\, \log\left(\frac{q_{y|x}}{q_y}\right)  \\\\
         &\displaystyle= \frac{1}{k_x}\sum_x \left[\log 2+q_{1|x}\log q_{1|x}+(1-q_{1|x})\log(1-q_{1|x})\right]=\frac{1}{k_x}\sum_x I_x(q_{1|x}).
    \end{array}
    \label{eq16:infodef}
\end{equation}
\end{linenomath*}
As in the previous example, the property $I(\bm{q})$ depends on the set of conditional probabilities $\bm{q}=\{q_{1|x}\}_{x=1}^{k_x}$.

The prior $p(\bm{q}|\beta)$ of Eq.~(\ref{eq07:priorMaxEnt}) reads
\begin{linenomath*}
\begin{equation}
    \displaystyle p(\bm{q}|\beta) \propto \exp\left[\beta I(\bm{q})\right] \propto \prod_x p(q_{1|x}|\beta).
    \label{eq17:priorInfo}
\end{equation}
\end{linenomath*}
Again, the prior factorises. The individual factors are
\begin{linenomath*}
\begin{equation}
    \displaystyle p(q_{1|x}|\beta) \propto \exp\left[\frac{\beta}{k_x}I_x(q_{1|x})\right]\propto \left[q_{1|x}^{q_{1|x}}\,(1-q_{1|x})^{(1-q_{1|x})}\right]^{\beta/k_x}.
    \label{eq18:indpriorInfo}
\end{equation}
\end{linenomath*}
This prior resembles the Beta distribution $\propto q_{1|x}^{\nu}\,(1-q_{1|x})^{\nu}$ proposed in \citep{hernandez2019estimating}, but has more complicated exponents. Each posterior reads
\begin{linenomath*}
\begin{equation}
    \displaystyle p(q_{1|x}|n_{1x},n_{0x},\beta)=\frac{ q_{1|x}^{n_{1x}}(1-q_{1|x})^{n_{0x}}\, e^{\beta I_x(q_{1|x})/k_x} }{\mathcal{Z}(n_{1x},n_{0x},\beta)}.
    \label{eq19:posteriorq1xInfo}
\end{equation}
\end{linenomath*}
This functional form does not allow us to continue the problem in a analytically tractable fashion. Yet, the factorisation turns the problem uni-dimensional, so it is easy to calculate the partition function for different pairs of $\{n_{1x},n_{0x}\}$ and the corresponding expectation values. Indeed, the partition function factorises as $\mathcal{Z}(\bm{n},\beta)=\prod_x \mathcal{Z}(n_{1x},n_{0x},\beta)$. 

\begin{figure}[H]
\includegraphics[width=12.5 cm]{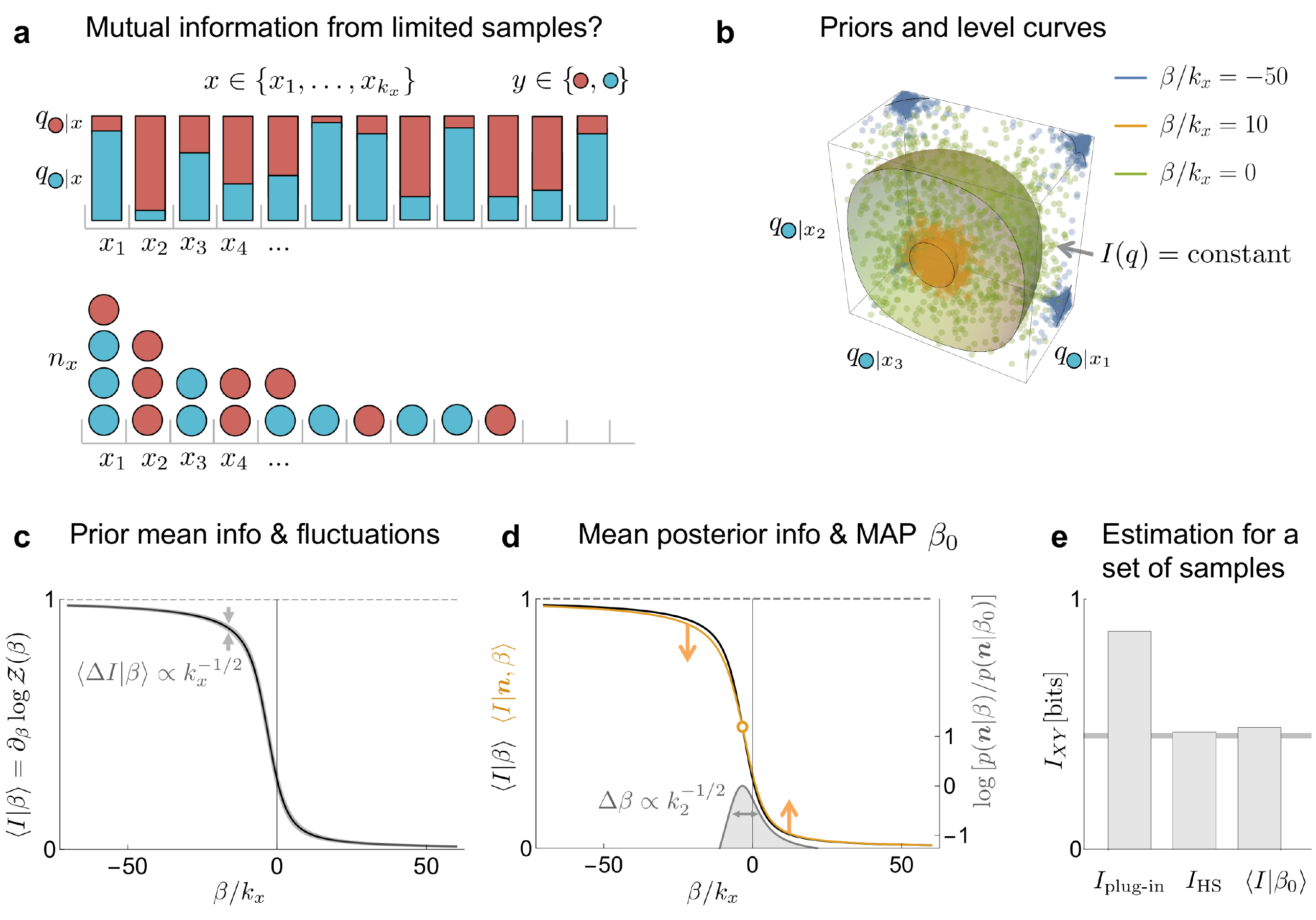}
\caption{Toy example used to illustrate the estimation of the amount of mutual information in the same system as in Sect.~\ref{sect:marginal}. (\textbf{a}) The system is governed by a bivariate probability distribution, with states characterised by two labels, $x$ and $y$. There are many $x$ states, $k_x\gg 1$, and all are equally probable, that is, $q_x=1/k_x$. The variable $y$ is binary, and its two values are depicted as red and blue, corresponding to $1$ and $0$, respectively. The conditional probabilities were sampled from a symmetric Beta distribution with parameter $0.5$. The goal is to estimate the mutual information $I_{XY}$ from $n$ samples, with $n \gtrsim k_x$. (\textbf{b}) The level surfaces of the mutual information, as well as some sampled $\bm{q}$-values are displayed in a three dimensional subspace of the full $\bm{q}$-space, for different values of the hyperparameter $\beta$. (\textbf{c}) Prior mean mutual information as a function of the scaled hyperparameter $\beta/k_x$, and its fluctuations for $k_x = 100$. (\textbf{d}) Prior and posterior mean mutual information as a function of the scaled hyperparameter $\beta/k_x$ for $n = 60$ samples. The intersection of these curves corresponds to the MAP estimation $\langle I|\beta_0\rangle$. In gray, the posterior marginal evidence for the hyperparameter, whose width decreases as the square root of the number of states with at least two samples, $k_2$. (\textbf{e}) Comparison of the estimation of mutual information between different methods for the considered set of samples ($I_{\text{HS}}$ is the estimator from \citep{hernandez2019estimating}). The horizontal line corresponds to the true value of the mutual information. \label{fig:fig_info}}
\end{figure}

The estimated value of the amount of mutual information of each $\beta$-value ---the member's guess--- can be obtained with Eq.~(\ref{eq09:memberPostGuess}), and the $\beta$-value that maximizes the marginal likelihood ---the member's evidence--- from Eq.~(\ref{eq11:peakMargLik}). Interestingly, not all $x$-states contribute to the marginal likelihood,
\begin{linenomath*}
\begin{equation}
    \log p(\bm{n}|\beta)=\log\left[ \frac{\mathcal{Z}(\bm{n},\beta)}{\mathcal{Z}_0(\beta)}\right]= \sum_x \log\left[ \frac{\mathcal{Z}(n_{1x},n_{0x},\beta)}{\mathcal{Z}(0,0,\beta)}\right]
    \label{eq20:marglikZx}
\end{equation}
\end{linenomath*}
since the only non-vanishing terms are those in which the partition function associated to the $x$ state is not proportional to the prior partition function, that is, $\mathcal{Z}(n_{1x},n_{0x},\beta) / \mathcal{Z}(0,0,\beta)$ varies with $\beta$. For mutual information, this condition is only met by $x$ states with coincidences ($n_x\ge 2$). In fact, only when coincidences begin to show up ($n \gtrsim \sqrt{k_x}$) the sampled data can start to provide evidence about a non-vanishing mutual information. No such condition was found in the previous section, since there, every sampled $x$-state contributed to the estimation of the marginal probability. This discrepancy stems from the fact that, while the marginal probability depends on the first moment of the set $\{q_{1|x}\}$ of conditional probabilities, the mutual information relates to the second moment, or the amount of dispersion, of this set \citep{hernandez2019estimating}. We show a toy example that illustrates this approach in Fig.~\ref{fig:fig_info}. In this example, the proposed method performs as well as some of the best known estimators (Fig.~\ref{fig:fig_info}e) designed for such undersampled regime.

\subsection{Entropy}

\label{sec:entropia}

Finally we consider the estimation of entropy, for which multiple estimators have been proposed in the past \citep{miller1955note, grassberger2003entropy, chao2003nonparametric, nemenman2004entropy, nemenman2011coincidences, chao2013entropy, berry2013simple, archer2014bayesian}. The entropy $H(\bm{q})$ is defined as
\begin{linenomath*}
\begin{equation}
    \displaystyle H(\bm{q})= -\sum_x q_x \log q_x.
    \label{eq21:entropydef}
\end{equation}
\end{linenomath*}
Here we use natural logarithm for analytical convenience. We assume that the number of accessible states is unknown, and that the number of effective states is quite large, namely $\exp[H(\bm{q})]\gg 1$.

\begin{figure}[H]
\includegraphics[width=12.5 cm]{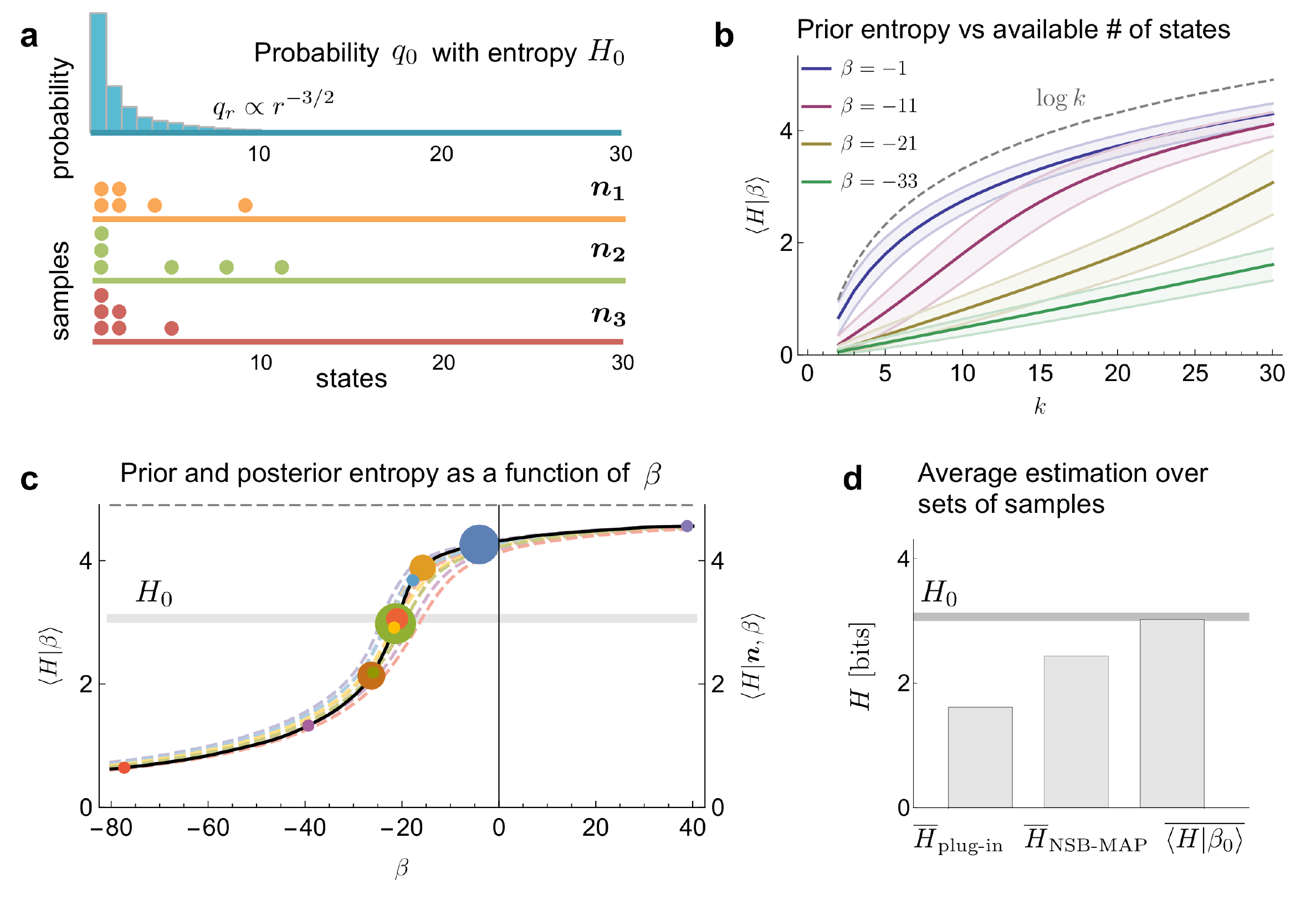}
\caption{Entropy estimation in a toy example. (\textbf{a}) A distribution $q_0$ has ranked probabilities that decrease with a power law ($q_r \propto r^{-3/2}$) within a finite number of states ($k = 30$). Distributions such as these represent a challenge for the estimation of entropy. Three possible sets of $n = 6$ samples are displayed. (\textbf{b}) Prior mean entropy, plus/minus a standard deviation, as a function of the available number of states $k$ for several values of the hyperparameter $\beta$. (\textbf{c}) Prior and posterior mean entropy as function of $\beta$ for the different possible sets of multiplicities that can be obtained from $
n = 6$ samples (three examples shown in panel a, with matching colors). The intersections between prior and posterior mean entropy (MAP estimation) are marked with circles. The size of the circles is proportional to the likelihood of the corresponding multiplicity (a minimun size is imposed, for visibility). (\textbf{d}) Comparison of the average estimation of entropy between different methods over all the multiplicity sets (discarding the set with no coincidences, and the set with all samples in one state), weighting according to their likelihood. The horizontal line corresponds to the true value of the entropy. \label{fig:fig_entrZ}}
\end{figure}

Once more, we expand the prior in a family whose members are tagged with $\beta$, and use Eq.~(\ref{eq07:priorMaxEnt}) to express each member of the family as
\begin{linenomath*}
\begin{equation}
    \displaystyle p(\bm{q}|\beta) \propto \exp\left[\beta H(\bm{q})\right] 
    = \prod_x q_x^{-\beta q_x},
    \label{eq22:priorqEntropy}
\end{equation}
\end{linenomath*}
where $\bm{q}$ belongs to the simplex, $\sum_x q_x=1$. We can compare this prior to the one used by NSB \citep{nemenman2004entropy}, a symmetric Dirichlet $p(\bm{q}|\beta) \propto \prod_x q_x^{\beta-1}$. Although there are some similarities, the prior in Eq.~(\ref{eq22:priorqEntropy}) has more complex exponents, which hamper the analytical tractability. The posterior reads
\begin{linenomath*}
\begin{equation}
    \displaystyle p(\bm{q}|\bm{n},\beta) = \frac{\delta\left(\sum_j q_j-1\right)\,\exp\left[\beta H(\bm{q})\right]\, \prod_x q_x^{n_x} }{\mathcal{Z}(\bm{n},\beta)}=\frac{\delta\left(\sum_j q_j-1\right) \prod_x q_x^{n_x-\beta q_x} }{\int {\rm d}\bm{q}'\, \delta\left(\sum_j q'_j-1\right) \prod_x {q'}_x^{(n_x-\beta q'_x)}}.
    \label{eq23:posteriorqEntropy}
\end{equation}
\end{linenomath*}
To calculate the member's estimation $\langle H | \bm{n}, \beta \rangle$ and the marginal likelihood $p(\bm{n}| \beta)$ the partition function $\mathcal{Z}(\bm{n},\beta)$ is needed.

We have found no evident way to solve the partition function analytically. Yet, its shape can be obtained by integrating numerically the partial convolutions that the delta over the simplex implies (for $k$), with which the prior and posterior estimates of the entropy, as well as the optimal $\beta$ could be obtained (Fig.~\ref{fig:fig_entrZ}c). The MAP estimator of the entropy obtained with this procedure, even if numerical, was more accurate than the results derived from all previous methods we know of (Fig.~\ref{fig:fig_entrZ}d).

\section{Discussion}

When inferring the value of properties $F(\bm{q})$, three sampling regimes are relevant. If the number of samples $n$ is significantly smaller than $\sqrt{\exp[H(\bm{q})]}$ the sample is likely to contain no coincidences \citep{nemenman2011coincidences}, and in this regime, no inference is possible. This study becomes useful when $\sqrt{\exp[H(\bm{q})]} \gtrsim n$, but still, the order of magnitude of $n$ is not significantly larger than $\exp[H(\bm{q})]$. For still larger $n$, the estimation of the probabilities $q_i$ begins to be feasible. 

The Bayesian approach is argued to be the only rational way to infer the value of an unknown property. Yet, the success of Bayesian inference relies of an adequate choice of the prior. If an unjustified prior is used, only because there are no evident reasons to choose a specific prior, the result of the inference is not guaranteed to make sense. It has been argued that more than a drawback of the Bayesian approach, this sensitivity to the prior stresses the fact that it is impossible to make inference without assumptions \citep{mackay2003}. Yet, it is often the case that selecting a prior is difficult, and this is problematic when the result is too sensitive to the choice, which is often the case in the severe undersampled regime. Without intending to diminish the relevance of priors, in this paper we have shown that not all aspects of the prior distribution are relevant, and thereby, we hope to focus the efforts of the selection process to the relevant aspects. For example, we proved that all the modulations of the prior within the intersection between a level surface of the property and a level surface of the likelihood bear no consequences on the result of the inference. This observation, and the understanding that the prior distribution cannot depend explicitly on the sampled data reduce the search of a high-dimensional prior $p(\bm{q})$ to a one-dimensional prior $g(f)$. Moreover, in many circumstances, even the detailed definition of $g(f)$ may be more than required, although this statement only holds when the sampled data contain a non-negligible fraction of coincidences. When this condition holds, the data themselves select a range of $f$-values that actually contribute to the estimation. Therefore, only the variations of the prior distribution $g(f)$ within this range actually matter. 

To make this selective role of the data explicit, and at the same time, to be able to get analytical results, in this paper we proposed to expand the prior distribution as a linear combination of base prior functions, each of which was required to have a different mean value of the property under study. By imposing no additional structure on the base priors (this requisite was instantiated by MaxEntropy), the distributions of the base were derived to depend exponentially on the property under study. The most relevant element of the base (at least, inasmuch the relevance was dictated by the data and not by $g(f)$) was shown to be the one for which the prior and posterior estimation of the property coincided (Eq.~\ref{eq11:peakMargLik}). This result is similar to the equality between the prior and posterior mean energy reported in the framework of statistical inference \citep{zdeborova2016}, and it relates to a scenario of parameter mismatch, as well as to the algorithm of expectation maximisation \citep{dempster1977maximum}. Even if the ratio of partition functions sometimes cannot be calculated analytically, the goal of this paper is to reveal the formal similarity between the two problems, and thereby, to prompt the reader to the battery of numerical algorithms that have been developed in expectation maximisation to solve the type of one-dimensional problem posed here.

In Sect.~\ref{sect:examples}, we tested the performance of the method for three well-known properties, for which many estimators  exist. Even though our method is general, it accomplished good results, when compared to those produced by other methods specifically designed for each individual property. Interestingly, the performance was good even using a single $\beta$-value, namely, the one for which the marginal likelihood $p(\beta|\mathbf{n})$ was maximal, without needing to integrate in $\beta$, and therefore, without needing to define the prior $g(\beta)$. One fortunate ingredient was given by the fact that in those examples, as well as in many cases of interest, the property to be estimated was additive and symmetric in the components $q_i$. Properties with these characteristics can be written as $F(\bm{q})=\sum_j \varphi(q_j)$, for some function $\varphi: \mathbb{R} \to \mathbb{R}$. When the effective number of states $\exp[H(\bm{q})]$ grows, the constraint imposed by the normalisation condition becomes increasingly irrelevant, so the sampled $q_i$ can be assumed to be fairly independent of one another. In this case, the central limit theorem implies that the relative fluctuations of the property $\sigma_{F|\bm{n}, \beta}/\langle F|\bm{n}, \beta \rangle$  decay as the inverse of the square root of the effective number of states. Therefore, each member of the base makes a rather definite  posterior guess $\langle F|\bm{n}, \beta \rangle$ with minimal fluctuations, implying that even though the states $p(\bm{q}|\beta)$ spreads all over the simplex, the posterior property is sharply defined within the state, as happened with the expansion in Delta-like priors.

In summary, the method proposed here has several attractive properties:
\begin{enumerate}
    \item[a)] It offers a general framework to select priors that is valid for arbitrary properties $F(\bm{q})$.
    \item[b)] It performs well in the estimation of well-known quantities, as marginal distributions, mutual information and entropy, comparable or even better than previous estimators individually tailored for each property.
    \item[c)] It reveals the formal similarity between inference problems and equilibrium statistical mechanics. In particular, the parameter $\beta$ is analogous to the inverse temperature of statistical physics, and the two alternative expansions of the prior (in a Delta-like or an exponential-like base) correspond to the microcanonical and the canonical ensembles, respectively.
\end{enumerate}




\authorcontributions{Conceptualization, DH and IS; methodology, DH and IS, software, DH; validation, DH; formal analysis, DH and IS; writing, DH and IS; visualization, DH.}

\funding{This research was funded by Universidad Nacional de Cuyo (Grant 06 / C589) and Agencia Nacional de Promoción Científica y Tecnológica (Grants PICT Raíces 2016 1004 and PICT 2019 2113).}

\institutionalreview{Not applicable.}

\informedconsent{Not applicable.}


\acknowledgments{The running costs of this study were covered by Consejo Nacional de Investigaciones Científicas y Técnicas and Comisión Nacional de Energía Atómica of Argentina.}

\conflictsofinterest{The authors declare no conflict of interest.'} 





\appendixtitles{no} 




\end{paracol}
\reftitle{References}


\externalbibliography{yes}
\bibliography{references}

\begin{thebibliography}{999}

\bibitem[Kraskov \em{et~al.}(2004)Kraskov, St{\"o}gbauer, and
  Grassberger]{kraskov2004estimating}
Kraskov, A.; St{\"o}gbauer, H.; Grassberger, P.
\newblock Estimating mutual information.
\newblock {\em Physical Review E} {\bf 2004}, {\em 69},~066138.

\bibitem[Nemenman \em{et~al.}(2004)Nemenman, Bialek, and van
  Steveninck]{nemenman2004entropy}
Nemenman, I.; Bialek, W.; van Steveninck, R.d.R.
\newblock Entropy and information in neural spike trains: Progress on the
  sampling problem.
\newblock {\em Physical Review E} {\bf 2004}, {\em 69},~056111.

\bibitem[Archer \em{et~al.}(2014)Archer, Park, and Pillow]{archer2014bayesian}
Archer, E.; Park, I.M.; Pillow, J.W.
\newblock Bayesian entropy estimation for countable discrete distributions.
\newblock {\em The Journal of Machine Learning Research} {\bf 2014}, {\em
  15},~2833--2868.

\bibitem[Chao \em{et~al.}(2013)Chao, Wang, and Jost]{chao2013entropy}
Chao, A.; Wang, Y.; Jost, L.
\newblock Entropy and the species accumulation curve: a novel entropy estimator
  via discovery rates of new species.
\newblock {\em Methods in Ecology and Evolution} {\bf 2013}, {\em
  4},~1091--1100.

\bibitem[Grassberger(2003)]{grassberger2003entropy}
Grassberger, P.
\newblock Entropy estimates from insufficient samplings.
\newblock {\em arXiv preprint physics/0307138} {\bf 2003}.

\bibitem[Archer \em{et~al.}(2013)Archer, Park, and Pillow]{archer2013bayesian}
Archer, E.; Park, I.M.; Pillow, J.W.
\newblock Bayesian and quasi-Bayesian estimators for mutual information from
  discrete data.
\newblock {\em Entropy} {\bf 2013}, {\em 15},~1738--1755.

\bibitem[Hern{\'a}ndez and Samengo(2019)]{hernandez2019estimating}
Hern{\'a}ndez, D.G.; Samengo, I.
\newblock Estimating the Mutual Information between Two Discrete, Asymmetric
  Variables with Limited Samples.
\newblock {\em Entropy} {\bf 2019}, {\em 21},~623.

\bibitem[Lehmann and Casella(1998)]{lehmann1998}
Lehmann, E.L.; Casella, G.
\newblock {\em Theory of Point Estimation}; Springer,  1998.

\bibitem[Cover and Thomas(2012)]{cover2012elements}
Cover, T.M.; Thomas, J.A.
\newblock {\em Elements of information theory}; John Wiley \& Sons,  2012.

\bibitem[Jaynes(1957{\natexlab{a}})]{jaynes1957a}
Jaynes, E.T.
\newblock Information Theory and Statistical Mechanics.
\newblock {\em Physical Review} {\bf 1957}, {\em 108},~171--190.

\bibitem[Jaynes(1957{\natexlab{b}})]{jaynes1957b}
Jaynes, E.T.
\newblock Information Theory and Statistical Mechanics II.
\newblock {\em Physical Review} {\bf 1957}, {\em 106},~620--630.

\bibitem[Amari and Hiroshi(2000)]{amari2000}
Amari, S.I.; Hiroshi, N.
\newblock {\em Methods of Information Geometry}; Oxford University Press,
  2000.

\bibitem[Amari(2001)]{amari2001}
Amari, S.I.
\newblock Information Geometry on Hierarchy of Probability Distributions.
\newblock {\em IEEE Transactions on Information Theory} {\bf 2001}, {\em
  47},~1701--1711.

\bibitem[Panzeri and Treves(1996)]{panzeri1996analytical}
Panzeri, S.; Treves, A.
\newblock Analytical estimates of limited sampling biases in different
  information measures.
\newblock {\em Network: Computation in neural systems} {\bf 1996}, {\em 7},~87.

\bibitem[Samengo(2002)]{samengo2002estimating}
Samengo, I.
\newblock Estimating probabilities from experimental frequencies.
\newblock {\em Physical Review E} {\bf 2002}, {\em 65},~046124.

\bibitem[Paninski(2003)]{paninski2003estimation}
Paninski, L.
\newblock Estimation of entropy and mutual information.
\newblock {\em Neural computation} {\bf 2003}, {\em 15},~1191--1253.

\bibitem[Abramowitz \em{et~al.}(1988)Abramowitz, Stegun, and
  Romer]{abramowitz1988handbook}
Abramowitz, M.; Stegun, I.A.; Romer, R.H.
\newblock Handbook of mathematical functions with formulas, graphs, and
  mathematical tables,  1988.

\bibitem[Montemurro \em{et~al.}(2007)Montemurro, Senatore, and
  Panzeri]{montemurro2007tight}
Montemurro, M.A.; Senatore, R.; Panzeri, S.
\newblock Tight data-robust bounds to mutual information combining shuffling
  and model selection techniques.
\newblock {\em Neural Computation} {\bf 2007}, {\em 19},~2913--2957.

\bibitem[Kolchinsky and Tracey(2017)]{kolchinsky2017estimating}
Kolchinsky, A.; Tracey, B.D.
\newblock Estimating mixture entropy with pairwise distances.
\newblock {\em Entropy} {\bf 2017}, {\em 19},~361.

\bibitem[Belghazi \em{et~al.}(2018)Belghazi, Baratin, Rajeshwar, Ozair, Bengio,
  Courville, and Hjelm]{belghazi2018mutual}
Belghazi, M.I.; Baratin, A.; Rajeshwar, S.; Ozair, S.; Bengio, Y.; Courville,
  A.; Hjelm, D.
\newblock Mutual information neural estimation.
\newblock  International Conference on Machine Learning. PMLR,  2018, pp.
  531--540.

\bibitem[Safaai \em{et~al.}(2018)Safaai, Onken, Harvey, and
  Panzeri]{safaai2018information}
Safaai, H.; Onken, A.; Harvey, C.D.; Panzeri, S.
\newblock Information estimation using nonparametric copulas.
\newblock {\em Physical Review E} {\bf 2018}, {\em 98},~053302.

\bibitem[Holmes and Nemenman(2019)]{holmes2019estimation}
Holmes, C.M.; Nemenman, I.
\newblock Estimation of mutual information for real-valued data with error bars
  and controlled bias.
\newblock {\em Physical Review E} {\bf 2019}, {\em 100},~022404.

\bibitem[Miller(1955)]{miller1955note}
Miller, G.
\newblock Note on the bias of information estimates.
\newblock {\em Information theory in psychology: Problems and methods} {\bf
  1955}.

\bibitem[Chao and Shen(2003)]{chao2003nonparametric}
Chao, A.; Shen, T.J.
\newblock Nonparametric estimation of Shannon’s index of diversity when there
  are unseen species in sample.
\newblock {\em Environmental and ecological statistics} {\bf 2003}, {\em
  10},~429--443.

\bibitem[Nemenman(2011)]{nemenman2011coincidences}
Nemenman, I.
\newblock Coincidences and estimation of entropies of random variables with
  large cardinalities.
\newblock {\em Entropy} {\bf 2011}, {\em 13},~2013--2023.

\bibitem[Berry~II \em{et~al.}(2013)Berry~II, Tka{\v{c}}ik, Dubuis, Marre, and
  da~Silveira]{berry2013simple}
Berry~II, M.J.; Tka{\v{c}}ik, G.; Dubuis, J.; Marre, O.; da~Silveira, R.A.
\newblock A simple method for estimating the entropy of neural activity.
\newblock {\em Journal of Statistical Mechanics: Theory and Experiment} {\bf
  2013}, {\em 2013},~P03015.

\bibitem[MacKay(2003)]{mackay2003}
MacKay, D.
\newblock {\em Information Theory, Inference and Learning Algorithms};
  Cambridge University Press,  2003.

\bibitem[Zdeborov\'a and Krzakala(2016)]{zdeborova2016}
Zdeborov\'a, L.; Krzakala, F.
\newblock Statistical physics of inference: thresholds and algorithms.
\newblock {\em Advances in Physics} {\bf 2016}, {\em 65},~453--552.

\bibitem[Dempster \em{et~al.}(1977)Dempster, Laird, and
  Rubin]{dempster1977maximum}
Dempster, A.P.; Laird, N.M.; Rubin, D.B.
\newblock Maximum likelihood from incomplete data via the EM algorithm.
\newblock {\em Journal of the Royal Statistical Society: Series B
  (Methodological)} {\bf 1977}, {\em 39},~1--22.

\end{thebibliography}

\end{document}